\documentclass[12pt]{article}
\usepackage{graphicx}
\usepackage{amssymb,amsmath,amsfonts,palatino,amsthm}
\usepackage{amssymb}
\usepackage{epstopdf}
\newcommand{\f}{\begin{equation}}
\newcommand{\ff}{\end{equation}}
\newcommand{\blankline}{\vskip .3cm}
\begin{document}

\title{Extending dualities to trialities deepens the foundations of dynamics}
\author{Lee Smolin\thanks{lsmolin@perimeterinstitute.ca} 
\\
\\
Perimeter Institute for Theoretical Physics,\\
31 Caroline Street North, Waterloo, Ontario N2J 2Y5, Canada\\
and\\
Department of Physics and Astronomy, University of Waterloo\\
and\\
Department of Philosophy, University of Toronto}
\date{\today}
\maketitle

\begin{abstract}
Dualities are often supposed to be foundational, but they may come into conflict with background independence, because a hidden fixed structures is needed to define the duality transformation.  This conflict can be eliminated by extending a duality to a triality.  This renders that fixed structure dynamical, while unifying it with the dual variables.  

To illustrate this, we study matrix models with a cubic action, and show how breaking its natural triality symmetry by imposing different compactifications  yields  particle mechanics, string theory and Chern-Simons theory.  These result from compactifying, respectively, one, two and three dimensions.  This may explain the origin of Born's duality between position and momenta operators in quantum theory, as well as some of the the dualities of string theory.

\end{abstract}

\newpage

\tableofcontents


\section{Introduction: raising dualities to trialities}

Dualities play a crucial role in framing the dynamics of physical theories.  Two examples out of many will suffice:

\begin{itemize}

\item{} The duality between configuration and momentum variables: $(x^\mu , p_\nu )$, expressed by the symplectic structure:
\f
S = \int d\tau p_\mu \frac{ d x^\mu}{d \tau} + \ldots
\ff

\item{}The duality between quantum state and dual state, or bra and ket, $|\Psi >$ and
$<\Phi |$, expressed by the inner product
\f
< \Phi | \Psi >
\ff
or, equivalently, by the Hermitian conjugate operation
\f
\dagger \cdot | \Psi > \rightarrow < \Psi |
\ff

\end{itemize}

String theory abounds in dualities such as $T$-duality.  And the $AdS-CFT$ duality appears to express a deep relationship between diffeomorphism invariance and conformal invariance\cite{AdS-SD}, which transcends its original expressions\cite{Maldacena}.

These and other examples lead us to ask if there might be a deeper principle which unifies these different dualities and explains their importance for fundamental physics.  In this note I would like to propose an answer to this question, which is that the common origin of the diverse dualities in physics is a deeper principle of triality.

To see how this arises, consider that fundamental physics has another imperative, just as important as the realization of dualities, which is background independence.  This principle asserts that the laws of physics not depend on any, fixed, non-dynamical background structures\cite{TN,3R,TR,SURT}.  Background independence underlies all our modern understanding of the fundamental forces as its manifestation in field theories is diffeomorphism invariance and local gauge invariance.

However, notice that background independence is in conflict with several of the dualities of physics because the expression of each duality relies on the presence of a third, silent, non-dynamical structure.

In the case of the classical duality between $x^\mu$ and $p_\mu$ there is the time derivative
$\frac{d}{d\tau}$.  So there really are { \it three }elements in that duality
\f  
x^\mu , \ \ \ p_\nu , \ \ \  \frac{d}{d\tau} .
\ff
which combine to make the symplectic form.  Of these, two,  $x^\mu$ and $p_\mu$, are dynamical, whilee the third, $\frac{d}{d\tau}$  , is non-dynamical.    If we believe in the imperative of background independence, we should make $\frac{d}{d\tau}$ dynamical, reinvisioning this duality as a triality.  A partial step towards this is to
extend $\frac{d}{d\tau}$ to a covariant derivative depending on a dynamical connection
\f
\frac{d}{d\tau} \rightarrow \frac{D}{d\tau} =\frac{d}{d\tau} + A.
\ff
We will see shortly how to complete this step, and put $x^\mu$, $p_\mu$ and
$\frac{D}{d\tau}$ on an equal footing.

Similarly, in the quantum state space duality there is a third non-dynamical structure, the Hermitian conjugate map
\f
\dagger : {\cal H} \rightarrow {\cal H}^*
\ff
Once again, there are really three actors in the game
\f
|\Psi >  ,   <\Phi | , \dagger
\ff 
the two of which are dynamical, the third of which is frozen, waiting to be brought to life.

Let us propose to resolve the tension between the two principles of background dependence and duality 
by the following idea:

\blankline
\blankline

{\bf Extend each duality to a triality by promoting the silent, non-dynamical structure needed to express a duality to a third, equal dynamical partner. }

\blankline
\blankline

To see how this would work, we imagine we code $x^\mu$, $p_\mu$ and $\frac{d}{d\tau}$ into three members of an algebra, 
$\cal A$, 
\f  
A = x^\mu , \ \ \ B= p_\nu , \ \ \  C= \frac{d}{d\tau} .
\ff
with a naturally defined triple product
\f
(A,B,C ) \in R
\ff
such that
\f
S = \int d\tau p_\mu \frac{d x^\mu}{d \tau} = (x^\mu , p_\mu , \frac{d}{d\tau} )
\ff
We propose a similar extension of quantum theory based on a triple product.  In this case we hypothesize the existence of an algebra, ${ \cal A}^q$, with an invariant triple product
\f
(A,B,C) 
\ff
such that, if we choose
\f
A =  | \Phi > <\Phi | , B = |\Psi ><\Psi |   ,  C=  \dagger
\ff 
we get
\f
(| \Psi >, | \Phi > , \dagger)= |<\Psi | \Phi >|^2
\ff
In the simplest case, the algebra ${ \cal A}^q$ is a matrix algebra and the fundamental formula for quantum probabilities is the trace,
\f
P= Tr \left ( A B C \right ) = A_a^b B_b^c C_c^a   .
\ff
If we take $A$ and $B$ to represent pure states
\f
A_a^b = \alpha_a \alpha^{\dagger b}, \ \ \ \  B_a^b = \beta_a \beta^{\dagger b},
\ff
and $C=I$ to represent the inner product, we have
\f
P= | <\alpha | \beta >|^2
\ff

One way to attempt the elevation of dualities to trialities is suggested by the following arguments:

\begin{itemize}

\item{} The duality between $p$ and $x$ is suggested by the form of the symplectic structure for particle mechanics, but that duality is only realizable in a non-relativistic context through the harmonic oscillator, i.e.
\f
S = \int d\tau p_\mu \dot{x}^\mu - N H (x,p)
\ff  
whereas the only form for the Hamiltonian, $H$ consistent with a symmetry between $x^\mu$ and 
$p_\mu$ is the harmonic oscillator,
\f
H = \frac{1}{2} ( p^2 + x^2  -E)
\ff
So at the level of particle dynamics there is a conflict between Born duality and relativistic invariance.
To resolve this and have both Born duality and relativity at the classical level we have to go to 
the string, as shown first by\cite{laurent-string},
\f
S^{string} = \int d\tau d\sigma \left [
p (\tau, \sigma )_\mu \dot{x}^\mu - N H - VD
\right ]
\ff
with
\f
H= \frac{1}{2} ( p^2 + (\partial_\sigma x)^2  -E)
\ff 
and 
\f
D= p_\mu \partial_\sigma x^\mu
\ff
This is symmetric under the interchange,
\f
p_\mu  \leftrightarrow \partial_\sigma x^\mu
\label{Born}
\ff
Furthermore, as explained in \cite{laurent-string}, in this context, Born duality (\ref{Born}) is the origin of $T$ duality.  

\item{} However, to make string theory background independent it seems necessary to extend to
a membrane theory.  This was the idea of realizing a background independent form of string theory through $\cal M$ theory\cite{m}.  Indeed there is a good case that different background independent string theories are unified in a membrane theory.

\item{}Indeed, the strategy of starting with a purely cubic background independent action,  whose solutions define a background, appeared earlier in string field theory\cite{SFT}.  These theories were, however, subject to technical issues, which I conjectured could be resolved by framing them as membrane theories 
rather than string theories\cite{mtheory,stringloop,EJA}.

\item{}Indeed, one way to express a membrane theory is through a matrix model\cite{CH,DHN,BFSS,IKKT,others,mtheory,stringloop,EJA}.  
However once one is studying non-linear dynamics for very large or infinite matrices there is a trick which can be used to reduce any non-linear dynamics to quadratic equations.  This is to reduce the degree of equations by introducing auxiliary fields-and then coding these auxiliary fields in the degrees of freedom of an expanded matrix.  You can do this to you reduce any non-linear equations to the simplest possible non-linear equations-which are quadratic equations.  Hence any matrix theory should be in its most unified and compressed form when described by a cubic action.   This suggests a theory based on an algebra with a triple product defining a generalized trace.  

\item{} Indeed in \cite{EJA} the triality of the octonions, which generates a symmetry of the exceptional Joran algebra,  is seen to extend the duality expressed by $9+1$ dimensional supersymmetry transformations.

\end{itemize}

These ideas are developed in the next section.   But, before going into details, I mention other, closely related ideas, which have been developed in past papers on cubic matrix models.  The cubic matrix models have been used to propose a unification of gravitational and Yang-Mills dynamics\cite{universal}.  An even deeper unification bringing together the law with the state is described in \cite{stateislaw}.  Finally,
quantum mechanics itself can be understood as a consequence of the matrix dynamics resulting from breakng the fundamental triality to a duality\cite{hiddenmatrix}.

\section{Cubic matrix theory as the template for dynamics}

To realize these ideas, we study an algebraic representation of a membrane theory, of the form previously 
studied in \cite{mtheory,stringloop,EJA,eteralee} whose degrees of freedom
are three elements of an algebra:  
\f
M_a \in {\cal A}
\ff
where $a=0,1,2$ and $\cal A$ is some algebra possessing a commutator and a trace.  We will require that
\f
Tr M_a =0
\label{tracefree}
\ff

The {\it fundamental action} is taken to be
\f
S^f = \epsilon^{abc} Tr M_a M_b M_c  
\label{Sf}
\ff
The equations of motion are
\f
\epsilon^{abc} [M_b , M_c ]  =0
\label{eom}
\ff

Let $\cal G$ be the automorphism group of $\cal A$.
Our action has a large gauge symmetry under:
\f
M_a \rightarrow g^{-1} M_a g, \ \ \ \ \ \ \  \ \ g \in {\cal G}.
\label{gauge}
\ff

If we are willing to introduce some background structure-in particular a $2+1$ dimensional fixed metric, $q_{ab}$, we  can extend the fundamental action by three terms.  We add a two component $SO(1,2)$ fermion,
$\Psi_{\alpha}$, $\alpha = 0,1$, whose components are also valued in $\cal A$ and add to the action
\f
S^\Psi = \tau^{a \alpha}_\beta Tr \bar{\Psi}^\beta [M_a , \Psi_\alpha ]
\ff
where $ \tau^{a \alpha}_\beta$ are the three Pauli matrices.  Here
\f
q^{ab} = \tau^{a \alpha}_\beta  \tau^{b \beta}_\alpha
\ff

This induces a mass like term
\f
S^m = \frac{m^2}{2} Tr M_a M_b q^{ab}
\label{mass}
\ff

as well as a matrix-Yang-Mills term
\f
S^{YM} = q^{ac} q^{bd} Tr [M_a , M_b ] [M_c , M_d ]  
\ff
Our new terms are also invariant under (\ref{gauge}), with
\f
\Psi_\alpha \rightarrow g^{-1} \Psi_\alpha g
\label{Psi-gauge}
\ff
The four actions are also invariant under the global $SO (1,2)$ symmetry group of $q_{ab}$:
\f
M_a \rightarrow \Lambda_a^b M_b, \ \ \ \ \Psi_\alpha \rightarrow U(\Lambda )_\alpha^\beta \Psi_\beta
\label{global}
\ff

Note that $S^f$ is independent of background structure, while $S^\Psi$, $S^m$ and $S^{YM}$ depend on a fixed choice of Pauli matrices and metric.  In \cite{universal} we seek to make these dynamical and part of $M_a$.

\subsection{Compactification}

We will make liberal use of the compactification trick\cite{taylor}.  First, pick 
${\cal A} = SU (N) \otimes {\cal A}^\prime$, where $N$ is very large and $ {\cal A}^\prime$ is some other algebra.  Then we can write 
\f
M_0 =\hat{\partial}_t I^\prime = \hat{\partial}_0 I^\prime
\ff
where the other matrices become functions of $t \in S^1 $ and
\f
[ \hat{\partial}_t ,    M_i ] = \frac{\partial M_i (t)}{\partial t}
\ff
where $i=1,2$.
We also have 
\f
Tr \rightarrow \int_{S^1} dt Tr^\prime
\ff

\subsection{The origin of Heisenberg mechanics}

We now show that the one dimensional compactification just described is the origin of the Heisenberg algebra.  Expanding around $M_0 = \hat{\partial}_t$ we write,
\f
M_0 = \hat{\partial}_t I^\prime + A_0 (t)
\ff
where we recall that 
\f
Tr^\prime A_0 =0
\label{tracefree2}
\ff

We can also rename the other two matrices
\f
M_1 = X (t), \ \ \ \ M_2 = P (t)
\ff
The fundamental action becomes
\f
S^f \rightarrow \int dt Tr^\prime P \frac{d X}{dt} + A_0 [X,P]
\label{SfH}
\ff
Remembering (\ref{tracefree2}), the equation of motion for $A_0$ yields
\f
[X,P] = \imath \hbar I^\prime
\ff
where $\hbar$ is an arbitrary constant 
of integration and the relation is presumed to hold in the large $N$ limit.

Unpacking this let us further decompose the algebra ${\cal A}^\prime$
\f
{\cal A}^\prime = {\cal T}^d \otimes {\cal H}^m
\ff
where ${\cal T}^d$ generates translations in $d$ dimensions and ${\cal H}^m$ are the Hermitian 
matrices in $m$ dimensions, in the limit $m \rightarrow \infty$.  (We could equivalently  take these latter to be hermitian observables algebra in some Hilbert space.)  Then we  can write
\f
M_1 = X^\mu (t) {\cal T}_\mu, \ \ \ \ M_2 = P_\mu (t) {\cal T}^\mu
\ff
where $X^\mu (t)$ and $ P_\mu (t)$ are each $d$ hermitian matrices.  
$ {\cal T}^\mu$ are translation generators which satisfy
\f
Tr  {\cal T}^\mu =0 , \ \ \ \ Tr  {\cal T}^\mu  {\cal T}^\nu = \eta^{\mu \nu}
\ff
Then the $A_0$ equations of motion become 
the Heisenberg algebra
\f
[X^\mu,P_\nu] = \imath \hbar \delta^\mu_\nu I
\ff

The action (\ref{SfH}) is invariant under Born duality:
\f
X \rightarrow P, \ \ \  \ \ P \rightarrow -X
\label{Born}
\ff
We  can see that this is a remnant of the $SO(1,2)$ symmetry (\ref{global}) of the Chern-simon matrix action $S^f$.   

We may note that the mass term, (\ref{mass}), leads to the Born invariant harmonic oscillator dynamics 
\f
S^m \rightarrow \int dt \frac{m^2}{2} Tr \left ( P^2 + X^2 + (\partial_0 + A_0)^2 \right )
\ff

\subsection{Recovery of the free relativistic particle}

Is there a way to get the action for a relativistic free particle from a cubic matrix theory?
There is at least one way which involves extending the comic action to a more general 
form, involving a single matrix $\bf M$, with action,
\f
S^{emf} = Tr {\bf M}^3
\ff
To reach the fundamental action (\ref{Sf}) we expand
\f
{\bf M} = M_a \tau^a
\ff
where $\tau^a$, $a=0,1,2 $ are the three Pauli matrices.  However, we  can expand to a fourth Pauli matrix 
\f
{\bf M} = M_\alpha \tau^\alpha
\ff
with $\alpha = a, 3$, with $\tau^3 = I^{2 \times 2}$.
We now, parametrize
\f
M_3 = {\cal N}
\ff
and introduce a scaling  parameter $\alpha$
\f
M_0 = \frac{2}{\alpha} \left ( \hat{\partial}_t I^\prime + A_0 (t) \right )
\ff
\f
M_1 = \alpha X^\mu (t) {\cal T}_\mu, \ \ \ \ M_2 = P_\mu (t) {\cal T}^\mu
\ff
We take the limit $\alpha \rightarrow 0$, which signals we are breaking Born duality, to find
\f
S^{emf} \rightarrow \int dt Tr^\prime \left ( P_\mu  \frac{d X^\mu }{dt} + A_0 [X^\mu,P_\mu] 
+ {\cal N} P_\mu P_\nu \eta^{\mu \nu}
\right )
\label{SemfH}
\ff

\subsection{The origin of string theory}

It seems that we cannot write an action for relativistic particle dynamics without violating Born duality.
To realize both Born duality and relativity we have to follow \cite{laurent-string} and extend from a particle to string theory.  Interestingly enough, the transition from particles to strings can be accomplished within the cubic matrix theory, by just changing the vacuum we compactify and expand around.  

To find string theory, we compactly on a two-torus
\f
M_0 = \hat{\partial}_\tau I^\prime + A_0 (\tau, \sigma), \ \ \ M_1 = \hat{\partial}_\sigma I^\prime + A_1 (\tau, \sigma), \ \ \ \ \   M_3 = X^\mu (\tau, \sigma) {\cal T}_\mu
\ff
$S^f$ yields
\f
S^{f}= \int_{T^2} d\tau d\sigma Tr f_{01} X + A_0 \partial_\sigma X - A_1 \partial_\tau X
\ff
By the trace free condition this vanishes.  i.e. $Tr ( A_{0,1} {\cal T}_\mu )=0$.
The induced term, $S^{YM}$ yields the string action coupled to a world sheet gauge theory\cite{matrixstring}.
\f
S^{YM}= \int_{T^2} d\tau d\sigma Tr^\prime \left [
f_{01}^2 + (\partial_\tau X^\mu    + [A_\tau , X^\mu ]  )^2 -  (\partial_\sigma X^\mu  + [A_\sigma , X^\mu ]   )^2
\right ]
\ff
where
\f
f_{0,1} = \partial_\tau A_1 - \partial_\sigma A_0 + [A_0, A_1 ] 
\ff

\subsection{Compactification to Chern-Simons theory}

Now we compactly on a three-torus, $T^3 = (S^1)^3$, and find a solution to the equations of motion
(\ref{eom})
\f
M_a = \hat{\partial}_a,  \ \ \ \  \  \  [ \hat{\partial}_a ,  \hat{\partial}_b ] =0
\ff
We expand around this solution
\f
M_a =  \hat{\partial}_a + A_a (x^a )
\ff
to find
\f
S^f \rightarrow S^{CS} = \int_{T^3} Tr^\prime A \wedge dA + A^3 
\ff

\section{Conclusion}

Inspired by some thoughts on eliminating background structure by extending dualities to trialities, we have seen how quantum particle mechanics, string theory and classical Chern-Simons theory (for $SU(m)$, in the limit $m \rightarrow \infty$) are all unified and explained as aspects of matrix Chern-Simons theory.
This is a theory based on a fundamental triality.  This illustrates our main claim, which is that the basic dynamics of relativistic particles and strings follows from breaking that basic triality, which holds at a background independent level, to a duality.  The breaking from a triality to a duality introduces background structures which allow the dynamics of string and particles to be defined and, by doing so, may give us insight into the role of dualities in the formulation of these dynamocal theories.

\section*{ACKNOWLEDGEMENTS}

These ideas first occurred to me during a bus ride in Santa Barbara in the summer of 1986, following conversations with Louis Crane, Gary Horowitz and Andy Strominger about purely cubic string field theories\cite{SFT}.  They inspired my work on cubic matrix 
models\cite{mtheory,stringloop,EJA}.  But my 
 interest in them was recently reinvigorated by several provocative 
 conversations with Laurent Freidel, about his work on incorporating relative locality into string theory\cite{laurent-string}.   

This research was supported in part by Perimeter Institute for Theoretical Physics. Research at Perimeter Institute is supported by the Government of Canada through Industry Canada and by the Province of Ontario through the Ministry of Research and Innovation. This research was also partly supported by grants from NSERC, FQXi and the John Templeton Foundation.

\end{document}